\documentclass[sigconf]{acmart}

\usepackage{booktabs} % For formal tables
\usepackage[toc,page]{appendix}
\usepackage{subcaption}
\usepackage[inline]{enumitem}
\usepackage{flushend}

\setcopyright{none}

\makeatletter
\newcommand{\inlineitem}[1][]{%
\ifnum\enit@type=\tw@
    {\descriptionlabel{#1}}
  \hspace{\labelsep}%
\else
  \ifnum\enit@type=\z@
       \refstepcounter{\@listctr}\fi
    \quad\@itemlabel\hspace{\labelsep}%
\fi}
\makeatother

\begin{document}

\title{Exploring the Ideal Depth of Neural Network when Predicting Question Deletion on Community Question Answering}

%\titlenote{Produces the permission block, and copyright information}

%\author{Anonymous Authors}

\author{Souvick Ghosh}
\affiliation{%
  \institution{SC\&I, Rutgers University}
  \streetaddress{4, Huntington Street,}
  \city{New Brunswick} 
  \state{NJ, USA} 
  \postcode{08854}
}
\email{souvick.ghosh@rutgers.edu}

\author{Satanu Ghosh}
\affiliation{%
  \institution{  Jadavpur University }
  \streetaddress{188, Raja S.C. Mallick Rd}
  \city{Kolkata} 
  \state{WB, India} 
  \postcode{700032}
}
\email{satanu.ghosh.94@gmail.com}

\renewcommand{\shortauthors}{Ghosh and Ghosh}

\renewcommand{\shorttitle}{Ideal Depth of NN in Question Deletion Prediction}

\begin{abstract}
In recent years, Community Question Answering (CQA) has emerged as a popular platform for knowledge curation and archival. An interesting aspect of question answering is that it combines aspects from natural language processing, information retrieval, and machine learning. In this paper, we have explored how the depth of the neural network influences the accuracy of prediction of deleted questions in question-answering forums. We have used different shallow and deep models for prediction and analyzed the relationships between number of hidden layers, accuracy, and computational time. The results suggest that while deep networks perform better than shallow networks in modeling complex non-linear functions, increasing the depth may not always produce desired results. We observe that the performance of the deep neural network suffers significantly due to vanishing gradients when large number of hidden layers are present. Constantly increasing the depth of the model increases accuracy initially, after which the accuracy plateaus, and finally drops. Adding each layer is also expensive in terms of the time required to train the model. This research is situated in the domain of neural information retrieval and contributes towards building a theory on how deep neural networks can be efficiently and accurately used for predicting question deletion. We predict deleted questions with more than 90\% accuracy using two to ten hidden layers, with less accurate results for shallower and deeper architectures.
\end{abstract}





\settopmatter{printacmref=false} % Removes citation information below abstract
\renewcommand\footnotetextcopyrightpermission[1]{} % removes footnote with conference information in first column
\pagestyle{plain} % removes running headers

\maketitle

% MAIN BODY OF THE PAPER

\section{Introduction}

%Traditional offline information seeking practices have relied on the interaction between the information seeker and the provider, through a series of dialogue which clarifies the information need of the former. This process of sense-making ~\cite{dervin1992mind} helps the seeker from overcoming his inadequacy in knowledge to make sense of a problematic situation. In online information seeking, the human mediator has been replaced by computers which act as interfaces between the user (by processing and reformulating the query) and the information sources. However, at the initial stages of the search session, humans often fail to articulate their information need ~\cite{taylor1967question} using keywords that are essential for retrieval. 

Question-answering forums, which may be open or closed domain, allow the users to share and construct knowledge collectively. The asker, aware of his anomalous state of knowledge ~\cite{belkin1980anomalous, belkin1982ask}, searches for his question in the forum. This initiates a human computer interaction where the system fetches the most relevant question which has already been asked and a ranked list of answers from the knowledge base. The two steps involved in this process are question-question similarity assessment and the answer relevance ranking. The ranking of the answers are often performed using community feedback and are therefore, more reliable than the search engine result pages.
If the answer is not present in the collection already, the asker posts the question in the forum and awaits responses. Through multiple rounds of interaction between the asker and the responder, the information need solidifies; the asker accepts one of the answers and is able to solve the problem. The use of natural language to frame the question helps in better communicating the problem, and leads to better answers. 
The motivation of the asker is not merely cognitive needs but also the possibility of social interactions with other users. A large number of question answering sites -- such as WikiAnswers\footnote{http://www.answers.com/Q/}, Quora\footnote{https://www.quora.com/}, Reddit\footnote{https://www.reddit.com/}, Stack Overflow\footnote{https://stackoverflow.com/}, and Brainly\footnote{https://brainly.co/} -- catering to different target audiences have gained massive popularity in the  last decade. WikiAnswers have around 137 million users while Reddit reported 234 million unique users in 2017. 

As the number of visitors have increased, the CQA forums have often been overwhelmed by the number of questions posted. Low quality and inappropriate questions need to be periodically deleted by the moderators (for e.g., Stack Overflow deleted 1.5 million posts in 2016) and this necessitates the development of algorithms to predict the deleted questions automatically. This can be done either before the user posts the question (allowing him to address the shortcomings) ~\cite{correa2014chaff}, or after the posting occurs (by filtering out questions for moderators). 

In this paper, we have explored how deep neural networks can be efficiently used for predicting deleted questions. Through our work, we aim to explore the feasibility of deep neural models in tackling information retrieval problems. We contribute towards the theory of neural IR by investigating how the depth of the neural network influences the accuracy of prediction of deleted questions. Keeping all other factors as constant, we have varied the number of hidden layers of the neural model to find the ideal fit for our problem. We show the relationships between accuracy of the model, the computational time, and the number of hidden layers.

The rest of the article is organized as follows: Section 2 discusses background and related works, while Section 3 presents the experimental methodology. The results have been explained in Section 4, and Section 5 concludes with the findings and future work.

\section{Background}

%Community Question Answering has drawn researchers from the areas of information science, machine learning, natural language processing, and social sciences. 
Adamic et al.~\cite{adamic2008knowledge} proposed that ``everyone knows something'' highlighting the wisdom of the crowd in collecting and organizing content in CQA sites ~\cite{wang2013wisdom}. 
% Other researches explored the motivations behind asking questions ~\cite{choi2016user, gazan2011social, wilson2000human} and the types of interactions between humans and the information systems ~\cite{belkin1980anomalous, kuhlthau1991inside, markey1981levels}. 
Question answering sites differ from web search applications in accessibility, implementation, and usage. In CQA forums, the user usually needs to register and follow community guidelines but is rewarded with more relevant and personalized answers. The response time is longer while the number of false positives are lower due to better relevance assessment.

~\citeauthor{choi2016user} ~\cite{choi2016user} classified question-answering sites into four categories - community-based, collaborative, social and expert. Such sites could be further categorized based on the domains - open or closed. %While domain specific knowledge and ontologies could be used in closed domain Q\&A sites (e.g. Stack Overflow), open domain sites (e.g. Reddit) offer a wide range of topics ranging from professional to personal and recreational. Researches in educational CQA provides insight into how question-answering can be modeled on educational theories to facilitate better learning experiences ~\cite{anderson2001taxonomy, krathwohl2002revision}. 
Researches have also focused on the quality of content in CQA sites ~\cite{ravi2014great, agichtein2008finding, wang2013wisdom, li2012analyzing, bian2009learning} and its correlation to user count. To maintain the content quality, some sites allow only expert users to answer the questions ~\cite{li2010routing, li2011question, zhou2012classification}, for free or fee, in topics of interest ~\cite{wang2013wisdom}. Others have used social networks to encourage community participation, or {\em gamification} to encourage and increase high quality content ~\cite{anderson2012facebook}.

Questions and posts are deleted for a variety of reasons -- vague and unclear, too difficult ~\cite{asaduzzaman2013answering}, ambiguous, inappropriate, repetitive or atypical ~\cite{rath2017discerning}. Automatic deletion of questions increases visibility of the popular questions, reduces maintenance cost, and increases user satisfaction. While most prediction models have relied on machine learning ~\cite{rath2017discerning, correa2014chaff, anderson2012facebook}, the recent successes of deep neural models in automatic speech recognition and computer vision calls for its application in solving problems in information retrieval. However, deep learning is still in a nascent stage where the engineering applications have preceded the development of theories. The solutions, often developed using trial and error, are borne out of researches in other domains, and ontologies are developed post factum.

The depth of the neural models is considered a critical factor in achieving highly accurate solutions ~\cite{hinton2006reducing, weston2012deep}. While some research works have  argued that increasing the depth does not always produce the desired results ~\cite{sun2016depth}, a majority of researchers believe that by increasing the number of hidden layers, accuracy can always be increased at the cost of computational time ~\cite{srivastava2015training, simonyan2014very, romero2014fitnets, he2015delving, szegedy2014scalable}. Telgarsky~\cite{telgarsky2016benefits} explained that if the number of nodes per layer, and the number of distinct parameters are kept constant, a deep model with $\Theta(k^3)$ layers can be approximated by a network with $\mathcal{O}(k)$ layers only if it has  $\Omega(2^k)$ nodes. Similarly, ~\citeauthor{bianchini2014complexity} ~\cite{bianchini2014complexity} and ~\citeauthor{montufar2014number} ~\cite{montufar2014number} suggested that representation of more complex functions needs very deep models. While it is hard to train shallow models with fixed units ~\cite{ba2014deep}, a recent study ~\cite{zagoruyko2016wide} shows how a 16-layer wide deep model can outperform a 1000-layer deep model. The major takeaway from all the researches is that deep learning is still in an exploratory stage where there is no generalized solution: problems need to be solved on a case-by-case basis.

We aim to contribute towards the development of theory in neural IR by exploring the importance of depth while predicting deleted questions. We use three factors for our analysis -- the number of layers in the model, the computational time, and the accuracy achieved in prediction.

\section{Experimental Methodology}

In our experiments, we have used a supervised approach to predict the deleted questions. 

\subsection {Data representation}

To answer our research question, we have collected 6,000 questions from a popular educational question answering site Brainly. Half of these questions were deleted by the forum moderators during content moderation. 

To predict the deleted questions, we obtain a dense representation of the question instances, each of which may contain one or many sentences. Each question is input to an embedding function $\mathcal{E}$, such that:
$\mathcal{E}: \mathcal{V} \rightarrow \mathbb{R}^{m}$ (where $\mathcal{V}$ denotes the vocabulary set and $m$ is the embedding dimension).

Next, we use Google's pre-trained Word2Vec\footnote{\url{https://radimrehurek.com/gensim/models/word2vec.html}} model which has a vocabulary of 3 million words obtained from the Google News dataset. An embedding dimension of 300 represents the output vectors. Each word in the question instance is plotted as a vector on a 300-dimensional space. The distance between any two points on the vector space is a measure of their semantic similarity \cite{mikolov2013distributed, mikolov2013efficient}. The word embeddings help in capturing the rich linguistic context of the words.
For a question q, which contain a total of $n$ words ($w$), the feature extraction function $\psi$ concatenates ($||$) the word embeddings of individual words (obtained using the embedding function) using the merge function M. The weak annotation $a_q$ for question q was appended at the end of the feature vector.

\begin{equation}
\psi(q) = [\mathcal{M}_{i=1}^n(\mathcal{E}(w_i) || a_q],
\end{equation}

The final representation was a vector of 72,001 dimensions. As the questions differ in the number of words that they contain, we use zero padding to convert each input vector to a fixed dimension.

\subsection{Deep Neural Model Architecture}

\begin{figure}[!htb]
	\centering
    \includegraphics[width=\linewidth, height = 0.2\paperheight]{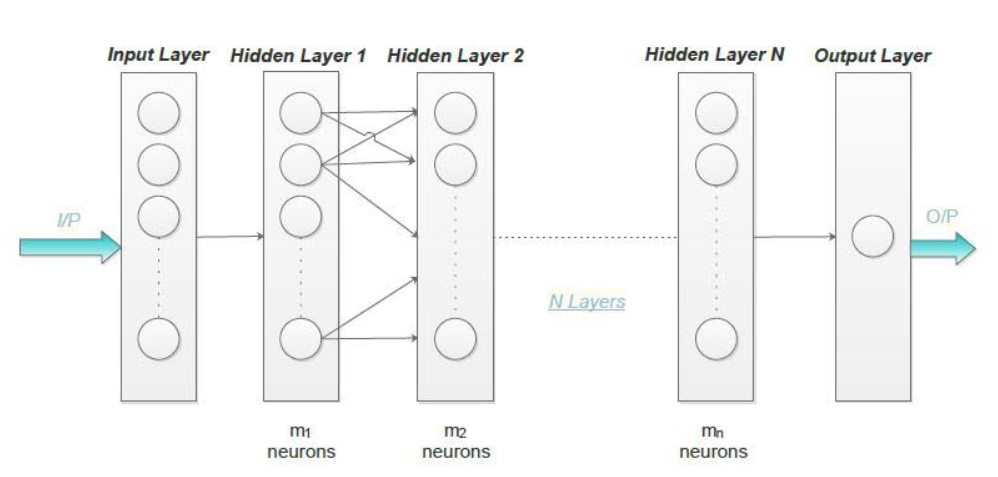}
    \caption{Deep Neural Model}
    \label{Fig:1-model architecture}
\end{figure}

To predict the deleted questions, we explore Multilayer Perceptron models with different number of hidden layers. Multilayer Perceptrons use backpropagation for supervised training and non-linear activation functions in the hidden layers.
The model architecture has been shown in Figure \ref{Fig:1-model architecture}. 
In our experiments, we have varied the number of hidden layers from 1 to 100. Each of the models have one input and one output layer and all the layers are densely connected. 
Hidden layers 1 to N have $m_1$, $m_2$, ..., $m_N$ neurons respectively where
$m_1$ $\geq$ $m_2$ $\geq$ ... $m_N-1$ $\geq$ $m_N$.
We have used Rectified Linear Units (popularly known as ReLU) in all the hidden layers. Sigmoid activation function is used in the output layer. We also apply a dropout of 5\% in all the hidden layers. As we predict only two categories, we use binary crossentropy as the loss function.

\section{Results}

For predicting question deletion, we have split our dataset into training and test sets. The training set contains 5,000 questions and the test set contains 1,000 questions. Both the training and test sets are balanced, containing equal number of deleted and non-deleted questions.

All the models are trained for 150 epochs and the results are reported in Table \ref{table:1 - results deletion}. The table shows the relationship between the number of hidden layers and the accuracy achieved on the training, the validation, and the test set. For validation, we have used 10\% of the training data in each iteration. The time required for training the model has also been reported in the table. For all the metrics, we have calculated the mean over multiple iterations.

\begin{table}[!htpb]
\small
\centering
\caption{\label{font-table} Influence of depth on prediction.}
\begin{tabular}{|p{1.25cm}|p{1.25cm}|p{1.25cm}|p{1.25cm}|p{1.25cm}|}
\hline
Number of Hidden Layers & Training Time (in secs) & Training Accuracy (\%) & Validation Accuracy (\%) & Test accuracy (\%) \\
\hline
1 & 6289 & 89.72 & 86.01 & 86.7 \\
2 & 6450 & 	94.58 & 91.01 & 90.4 \\
3 & 6623 & 99.39 & 95.81 & 96.2 \\
5 & 6644 & 99.80 & 98.00 & 97.5 \\
10 & 7056 & 99.81 & 98.40 & 97.7 \\
25 & 8399 & 99.29 & 97.21 & 96.6 \\
50 & 10549 & 59.33 & 52.10 & 50.4 \\
100 & 15644 & 59.63 & 47.21 & 49.2 \\
\hline
\end{tabular}
\label{table:1 - results deletion}
\end{table}

\begin{figure}[!htb]
	\centering
	\begin{subfigure}[t]{0.5\linewidth}
		\centering
        \includegraphics[width=\linewidth, height = 0.145\paperheight]{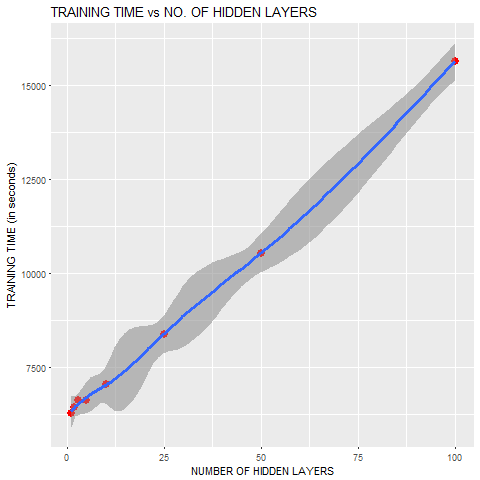}
        \caption{\label{font-figure} Training Time vs Depth.}
        \label{fig:2a}
	\end{subfigure}%
    ~
    \begin{subfigure}[t]{0.5\linewidth}
		\centering
        \includegraphics[width=\linewidth, height = 0.145\paperheight]{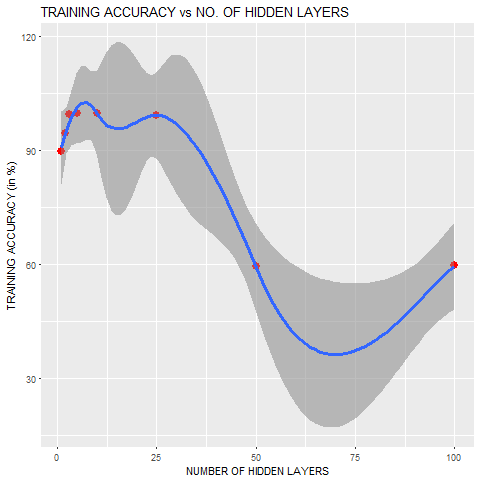}
        \caption{\label{font-figure} Training Accuracy vs Depth.}
        \label{fig:2b}
	\end{subfigure}
    
    \begin{subfigure}[t]{0.5\linewidth}
		\centering
        \includegraphics[width=\linewidth, height = 0.145\paperheight]{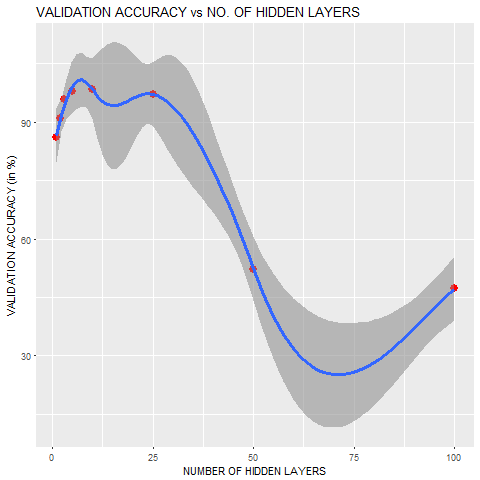}
        \caption{\label{font-figure} Validation Accuracy vs Depth.}
        \label{fig:2c}
	\end{subfigure}%
    ~
    \begin{subfigure}[t]{0.5\linewidth}
		\centering
        \includegraphics[width=\linewidth, height = 0.145\paperheight]{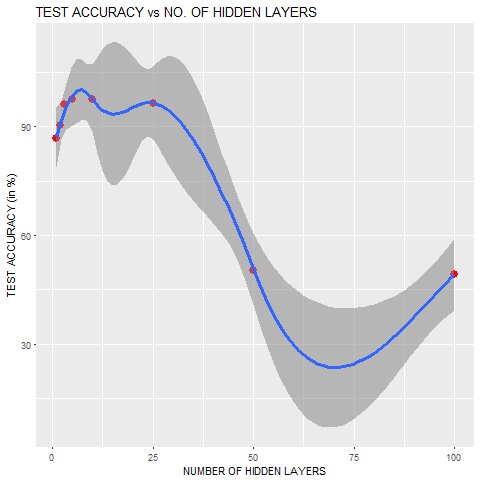}
        \caption{\label{font-figure} Test Accuracy vs Depth.}
        \label{fig:2d}
	\end{subfigure}
    \caption{Variation of Performance Metrics with Depth}
    \label{Fig:2-depth}
\end{figure}

In Figure \ref{Fig:2-depth}, we have plotted the curve of different performance metrics -- training time, training accuracy, validation accuracy, and test accuracy -- against the number of hidden layers (depth) in the neural network model. From Figure \ref{fig:2a}, it is evident that the training time of the model increases almost linearly with the number of hidden layers. Figures \ref{fig:2b}, \ref{fig:2c} and \ref{fig:2d} highlight that the accuracy of training, validation, and test increases as the number of hidden layers increase from one to two and from two to three. However, there is no significant increase in accuracy thereafter. The accuracy drops slightly from 10 to 25 hidden layers, but more significant drops are observed as we keep increasing the number of layers further.
The results highlight that there is no single deep neural architecture which solves all the problems. Depending on the research question, the researchers must develop the best architecture which is accurate and computationally feasible. As the backpropagation algorithm calculates derivatives using chain rule, higher number of layers often lead to the vanishing gradient problem. For predicting question deletion, the best results were obtained using two to five hidden layers, with each layer causing marginal increase in accuracy at the expense of training time.

\section{Conclusion}
In this paper, we explored how the depth of the neural network influences the accuracy of prediction of deleted questions in question-answering forums. We used different shallow and deep models for prediction and analyzed the relationships between the number of hidden layers, the accuracy of prediction, and the computational time. The results suggest that while deep networks perform better than shallow networks in modeling complex non-linear functions, increasing the depth may not always produce desired results. We observe that the performance of the deep neural model suffers significantly due to vanishing gradients when large number of hidden layers are present. Constantly increasing the depth of the model increases accuracy initially, after which the accuracy plateaus, and finally drops. Adding each layer is also expensive in terms of the time required to train the model. 

This research contributes towards building a theory on how deep neural networks can efficiently and accurately be used to predict question deletion in community Q\&A (CQA) forums. We predict deleted questions with more than 90\% accuracy using two to ten hidden layers, with less accurate results for shallower and deeper architectures. The results confirm that problems in neural IR need much exploration and there is no single solution to all the problems. While the number of hidden layers in critical in obtaining high accuracy, the computational cost incurred must also be considered. Once the accuracy reaches a saturation point, other parameters in the neural model should be altered to check for higher accuracy. In future, we would like to investigate similar problems in neural IR to obtain accurate and scalable architectures.

%\begin{acks}
%[Removed for anonymous reviewing.]
%We would like to thank Arvind Yelavarti, Manasa Rath, and Soumik Mondal who made significant contributions in data collection. Arvind also helped with feature extraction and integrity checking of the collected data. We would also like to thank Brainly who allowed us access to their questions database.
%\end{acks}

% BIBLIOGRAPHY 
\bibliographystyle{ACM-Reference-Format}
\bibliography{references} 

\begin{appendices}

\section{Additional Tables}

\begin{table}[h]
\small
\centering
\caption{\label{font-table} Examples of Questions Deleted and Not Deleted}
\begin{tabular}{|p{2cm}|p{5.25cm}|}
\hline
\textit{\textbf{Deleted}} & \textit{\textbf{Not Deleted}} \\ 
\hline
hello emo family. :3 wassup? & The sum of the angle measures of any triangle is 180°. Find each of the angle measures of a triangle if the second angle measures 10° more than twice the first, and the third angle measures 10° more than the second.\\ 
\hline
\end{tabular}
\end{table}

\begin{table}[h]
\small
\centering
\caption{\label{font-table} Stored Attributes in QA Forum Database}
\begin{tabular}{|p{2.25 cm}|p{5cm}|}
\hline
\textit{\textbf{Attributes}} & \textit{\textbf{Description}} \\ \hline
question\_id & Unique identifier of the question                    \\
user\_id     & Unique identifier for every user                     \\
subject\_id  & Unique identifier pertaining to the subject/domain   \\          client\_type\_id & Identifier for the type of client recorded in the forum \\
user\_category\_id  &  Category of the user\\
date\_created &  The date when the question was created     \\
date\_edited & The date when the question was last edited \\
date\_deleted  &  The date when the question was deleted \\
content &  The question with all its text\\
deleted\_type & The moderator annotated reason for deletion \\
\hline
\end{tabular}
\end{table}

\end{appendices}

\end{document}